\documentclass[aps,prb,twocolumn]{revtex4}
\usepackage{amsfonts}
\usepackage{amsmath}
\usepackage{amssymb}
\usepackage{graphicx}%
\begin{document}
\title{Symmetry-allowed phase transitions realized by the two-dimensional fully frustrated XY class}
\author{Petter Minnhagen$^{1}$}
\author{Beom Jun Kim$^{2}$}
\author{Sebastian Bernhardsson$^{1}$}
\author{Gerardo Cristofano$^{3}$}
\affiliation{$^{1}$Dept.\ of Physics, Ume{\aa } Univ., 901 87 Ume{\aa }, Sweden,
$^{2}$Dept.\ of Physics, Sungkyunkwan Univ., Suwon 440-746, Korea,$^{3}$Dept.\ of Physics,
Univ.\ of Naples, 80126 Naples, Italy}

\begin{abstract}
A 2D Fully Frustrated XY(FFXY) class of models is shown to
contain a new groundstate in addition to the checkerboard groundstates of the
standard 2D FFXY model. The spin configuration of this additional groundstate is obtained. Associated with this groundstate there are additional  phase transitions. An order parameter accounting for these new transitions is proposed. The transitions associated with the new order parameter are suggested to be similar to a 2D liquid-gas transition which implies $Z_{2}$-Ising like transitions. This suggests that the class of 2D FFXY models belongs within a $U(1)\otimes
Z_{2}\otimes Z_{2}$-designation of possible transitions, which implies that there are seven
different possible single and combined transitions.
MC-simulations for the generalized fully frustrated XY (GFFXY) model on a
square lattice are used to investigate which of these possibilities can be
realized in practice: five of the seven are encountered. Four critical points are deduced from the MC-simulations,
three consistent with central charge $c=3/2$ and one with $c=1$. The implications for the standard 2D FFXY-model are discussed in particular with respect to the long standing controversy
concerning the characteristics of its phase transitions.
\end{abstract}

\maketitle

\section{Introduction}

The two-dimensional(2D) fully frustrated XY (FFXY)-model describes a 2D 
Josephson junction array in a perpendicular magnetic field with the strength of the
magnetic field corresponding to one magnetic flux quanta for every second
plaquette of the array. The phase transitions of this model on a square lattice have been the subject of a long
controversy\cite{hasenbusch05r}$^-$\cite{korshunov02}.%\cite{hasenbusch05r}\cite{granato91}\cite{lee91b}\cite{lee91}\cite{granato93}\cite{olsson95}\cite{boubcheur98}\cite{korshunov02}
The emerging canonical picture is that the
model has two relevant phase ordering symmetries: an angular $\mathcal{U}(1)$-symmetry and a
$Z_{2}$-chirality symmetry \cite{villain77}$^,$\cite{teitel83}$^,$\cite{halsey85}. As a
consequence, the model has often been assumed to belong within the designation
$\mathcal{U}(1)\otimes Z_{2}$ \cite{granato91}$^,$\cite{lee91b}$^,$\cite{lee91}$^,$\cite{nightingale95}. 
The controversial questions have been: Does the model
undergo a single combined transition or two separate transitions and, if the
latter, in which order do the transitions occur? The emerging consensus is
two separate transitions: as the temperature is increased first a
Kosterlitz-Thouless(KT) transition associated with the angular $\mathcal{U}(1)$-
symmetry and then at a slightly higher temperature a $Z_{2}$-chirality
transition \cite{hasenbusch05r}$^,$\cite{korshunov02}. The cause of the controversy can, retrospectively, be attributed to the fact that the two transitions are extremely close in temperature.

 We here generalize the 2D FFXY model into a a wider 2D FFXY-class of models by changing the nearest neighbor interaction in such a way as to keep 
all symmetries. This generalized 2D FFXY-class is shown to contain an additional groundstate. The existence of this additional grounstate leads to a phase diagram containing four sectors.\cite{minnhagen07}. We here show that it has seven different phase transitions lines and four multicritical points. We use Monte Carlo simulations to establish the characters of the transitions of this phase diagram. Our simulations suggest that three of the critical points are consistent with the central charge $c=3/2$ and one with $c=1$.

In section 2 we define the 2D FFXY-model and in section 3 we describe the
structure of the new ground state. In section 4 we propose an order
parameter associated with the transition into this new
groundstate. In section 5 we give the results for the various phase transitions obtained from Monte Carlo simulations and determine the character of the four multicritical points by invoking a relation between the central charge $c$ and the bulk critical indices. In section 6 we discuss the original 2D FFXY\ model in view of our results. We also comment on related models not contained within the class of fully frustrated XY\ model discussed in the present investigation. Finally, some concluding remarks are given in section 7.

\section{Generalized Fully Frustrated XY Model}

The Hamiltonian which defines the 2D fully frustrated XY-class models on an
$L\times L$ square lattice is given by

\begin{equation}
H=\sum_{\langle ij\rangle}U\left(\phi_{ij}\equiv\theta_{i}-\theta_{j}-A_{ij}\right),
\end{equation}

with $\phi_{ij}\in\lbrack-\pi,\pi]$, where the sum is over nearest
neighbor pairs. The phase angle $\theta_{i}$ for the $i$th site at the lattice
point $(x_{i},y_{i})$ satisfies the periodicity $\theta_{i+L\hat{\mathbf{x}}%
}=\theta_{i+L\hat{\mathbf{y}}}=\theta_{i}$. The magnetic bond angle $A_{ij}$
is defined as the line integral along the link from $i$ to $j$, i.e.\
$A_{ij}\equiv(2\pi/\Phi_{0})\int_{i}^{j}\mathbf{A}\cdot d\mathbf{l}$ with the
magnetic vector potential $\mathbf{A}$ for the uniform magnetic field
$\mathbf{B}=B_{0}{\hat{z}}$ in the $z$ direction. With the Landau gauge taken,
$A_{ij}=2\pi fx_{i}$ for the vertical link and $A_{ij}=0$ for the horizontal
one, where the frustration parameter $f$ measures the average number of flux
quanta per plaquette. The fully frustrated
case corresponds to $f=1/2$ with a half flux quantum per plaquette on the
average. The Boltzmann factor, which determines the thermodynamic properties,
is given by $\exp(-H/T)$ where $T$ is the temperature. The interaction
potential $U(\phi)=U(\phi\pm2\pi)$ is periodic in $2\pi$ and is quadratic to
lowest order in $\phi$ so that $U(\phi)\sim\phi^{2}$. These conditions for
the interaction potential defines the class: the members of this class
are distinguished by the explicit form of the interaction potential
$U(\phi)$. If the relevant symmetry class is $\mathcal{U}(1)\otimes Z_{2}$, then in
principle three transitions are possible: separate $\mathcal{U}(1)$ and $Z_2$ transitions or a
merged $\mathcal{U}(1)\otimes Z_{2}$ transition. However, the number of allowed
phase transitions for the FFXY-class
is much larger\cite{minnhagen07}. \emph{The
implication is that by just changing the specific form of $U(\phi)$ within the
FFXY-class one could encounter a plethora of phase transitions}. In order to
verify this, we choose a parametrization of $U(\phi)$ and find the phase
transitions corresponding to this parametrization using Monte Carlo
simulations techniques. This strategy was employed earlier in
Ref.\ \cite{minnhagen07}. The parametrization
is of the form $U(\phi)$ where \cite{domani84}$^,$\cite{jonsson94}

\begin{equation}
U(\phi)=\frac{2}{p^{2}}[1-\cos^{2p^{2}}(\frac{\phi}{2})]\label{para}%
\end{equation}

and $p=1$ corresponds to the standard FFXY since $2[1-\cos^{2}(\phi
/2))]=1-\cos(\phi)$. The members of the FFXY class, which belong to this
parametrization, was in Ref.\ \cite{minnhagen07} termed the Generalized Fully
Frustrated XY (GFFXY) model. Figure 1a shows a sequence of interaction potentials $U(\phi)$.

\begin{figure}[ptb]
\begin{center}
\includegraphics[width=\columnwidth]{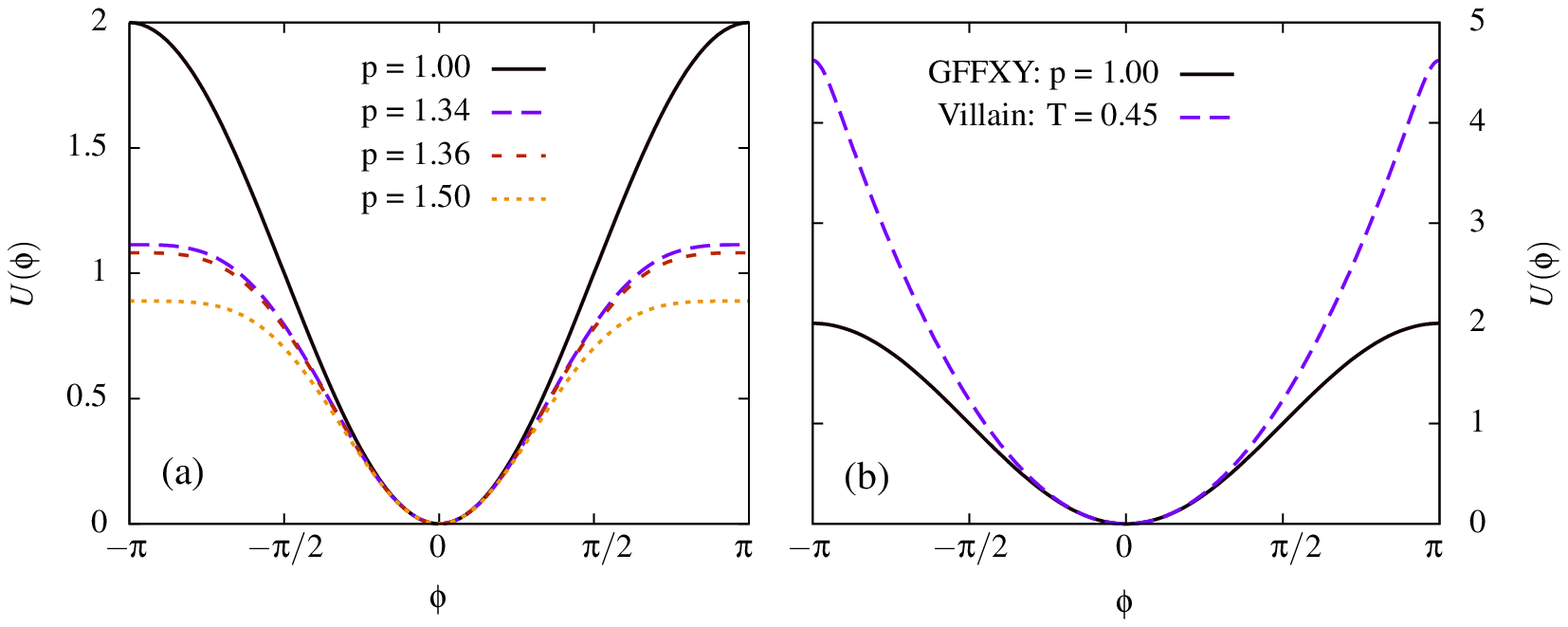}
\caption{Interaction potentials $U(\phi)$ in Eq.\ 2 at various values of $p$
are compared in (a). The standard XY model corresponding to $p=1$
is also compared with the Villain interaction potential in (b).
All interactions have the same symmetry and have the identical quadratic
form at small $\phi$.}
\end{center}
\end{figure}

To sum up:  The 2D FFXY class which we discuss here is obtained from the standard 2D FFXY by generalizing the interaction potential within the allowed conditions: $U(\phi)$ is a monotonously increasing function in the interval $\phi\in[0,\pi]$, $U(\phi)=U(\phi\pm2\pi)$ is periodic in $2\pi$ and is quadratic to lowest order in $\phi$ so that $U(\phi)\sim\phi^{2}$. The GFFXY model is by construction contained within this class. The Villain interaction is also contained in this class\cite{villain77}. In Fig.\ 1b the interaction potential for the standard XY model
$U(\phi)=1-\cos(\phi)$ is compared to the one for the Villain model at the
KT-transition ($T=0.45$) $U(\phi)=-T\ln\{\sum_{n=-\infty}^{n=\infty}%
\exp(-(\phi-2\pi n)^{2}/2T)\}$.\cite{villain77}$^,$\cite{hasenbusch05r}. The 2D
FFXY model with the Villain interaction has the same phase transition scenario
as the usual 2D FFXY model i.e. a $\mathcal{U}(1)$ KT-transition followed
by a $Z_{2}$ transition (still extremely close together but a little less
close than for the standard 2D FFXY model).\cite{hasenbusch05r} Is this true
for all models within the FFXY class? The answer is no.\cite{minnhagen07} The
reason is, according to us, connected to the appearance of a new groundstate.

\section{Groundstate}

\begin{figure}[ptb]
\begin{center}
\includegraphics[width=\columnwidth]{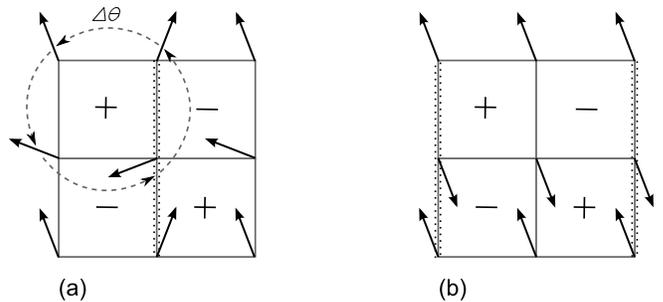}
\caption{Two groups of distinct groundstates of the 2D GFFXY model.
(a) When $p$ is smaller than $p_c (\approx 1.3479)$, the gauge-invariant
phase difference $\phi = \pi/4$ for all edges of a plaquette.
(b) When $p > p_c$, one edge has $\phi = \pi$ while all other three
have $\phi = 0$. The wiggled vertical lines denote the magnetic
bond angles $A_{ij} = \pi$, arrows indicate phase values, and $\pm$
represent vortex charges.}
\end{center}
\end{figure}

Let us first consider the groundstate for the standard 2D FFXY model on a
square lattice: The spin configuration corresponding to the groundstate
checkerboard is given in Fig.\ 2a.\cite{halsey85} A square with (without) a flux
quanta is denoted by $+$ ($-$). The arrows give the spin directions and the
thick (thin) links are the links with (without) magnetic bond angles $\pi$ ($0$) modulo
$2\pi$. In this configuration all the links contribute the same energy
$U(\frac{\pi}{4})$ to the groundstate. Thus the energy for the four links
constituting an elementary square is in this configuration $4U(\frac{\pi}{4})$.
The broken symmetry of the free energy is for $T=0$ directly related to the
fact that in order to change $+$ to $-$ squares in Fig.\ 2a by continuously
turning the spin directions from the one groundstate to the other, an
increase of the energy is required by a finite amount for a number of links. This required number of
links goes to infinity with the size of the system: the two groundstates are
separated by an infinite energy barrier in the thermodynamic limit.

The crucial point in the present context is that the groundstate shown in
Fig.\ 2a does not remain the groundstate for all values $p$. As $p$ is increased,
the maximum link energy $U(\pi)$ decreases and at a particular value $p_{c}>1$
the groundstate switches to the spin configuration shown in Fig.\ 2b. The energy
for the links around a square is for this configuration given by
$U(\pi)+3U(0)$. The critical value $p_{c}$ is hence given by the condition
$U(\pi)+3U(0)=4U(\frac{\pi}{4})$ leading to the determination%

\begin{equation}
p_{c}=\sqrt{\frac{\ln(3/4)}{2\ln(\cos(\pi/8)}}=1.3479.
\end{equation}

The groundstate for $p>p_{c}$ shown in Fig.\ 2b has the property that
an infinitesimal change of the middle spin is enough to flip between the two
checkerboard patterns (switching between $+$ and $-$ in Fig.\ 2b). Thus there is
no barrier between these two checkerboard patterns for $p>p_{c}$. This means
that the broken symmetry of the free energy associated with the two possible
checkerboard patterns states is restored. However, there is a new infinite
barrier between the two degenerate groundstates on opposite sides of $p_{c}$:
continuously turning the spins to change from the spin-configuration in Fig.\ 2a
to the spin-configuration in Fig.\ 2b requires an infinite energy.

\section{Order Parameters}

In order to characterize the phase transition properties of the 2D
GFFXY model one needs to identify a set of order parameters with which all possible transitions can be characterized:

\begin{figure}[ptb]
\begin{center}
\includegraphics[width=\columnwidth]{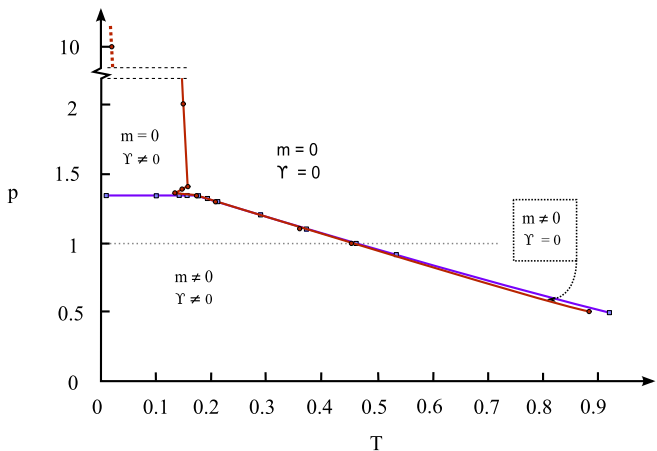}
\caption{Phase diagram of the 2D GFFXY model in the $(p,T)$ plane. The staggered
magnetization $m$ and the helicity modulus $Y$ give us all four
combinations, all of which are realized in the phase diagram. The horizontal
dotted line at $p=1$ corresponds to the standard FFXY model which has two
distinct, extremely close transitions.}
\end{center}
\end{figure}

The checkerboard pattern is usually associated with a $Z_{2}$ chirality symmetry. For $T=0$ this symmetry is reflected in the existence of two degenerate groundstates (the two checkerboards) separated by an infinite energy barrier.
The corresponding order parameter is related to the staggered magnetization $m$ defined
as~\cite{teitel83}

\begin{equation}
m=\left\langle \left\vert \frac{1}{L^{2}}\sum_{l=1}^{L^{2}}(-1)^{x_{l}+y_{l}%
}s_{l}\right\vert \right\rangle ,
\end{equation}

where $\langle\cdots\rangle$ is the ensemble average and the vorticity for the
$l$th elementary plaquette at $(x_{l},y_{l})$ is computed from $s_{l}%
\equiv(1/\pi)\sum_{\langle ij\rangle\in l}\phi_{ij}=\pm1$ with the sum taken
anti-clockwise around the given plaquette. The corresponding broken symmetry is reflected 
in the following way: for any finite system the quantity
$\frac{1}{L^{2}}\sum_{l=1}^{L^{2}}(-1)^{x_{l}+y_{l}}s_{l}$
can with finite probability acquire any value in the range $[-1,1]$ allowed by
the model. However, in the thermodynamic limit $L=\infty$ only values in
either the range $[-1,0]$ or the range $[0,1]$ can be acquired. This means that the order parameter $\mathcal{O}=\langle \frac{1}{L^{2}}\sum_{l=1}^{L^{2}}(-1)^{x_{l}+y_{l}}s_{l}\rangle$ in the tthermodynamiclimit only can take on the two values $\mathcal{O}=\pm m$. The probability for the two values are equal but they are separated by an infinite free-energy barrier. This is equivalent to saying that the order parameter $\mathcal{O}$ has a $Z_{2}$ symmetry which is broken. In the broken symmetry region  $m\neq0$ whereas when the symmetry is unbroken $m=0$. Figure 3 shows the phase diagram in the $(T,p)$-plane. As seen $m\neq0$ corresponds to a finite region of this plane.

For $T=0$ and $p=p_{c}$, the two groundstates in Fig.\ 2a and b are degenerate and separated by an infinite energy barrier. For $T>0$ this should instead take the form of an infinite free energy barrier in the thermodynamic limit, separating values which a local order parameter can acquire. To this end one needs to identify an appropriate local order parameter. Such a possible order parameter is the defect density $n_{k}$ defined
by

\begin{equation}
n_{k}=\left\langle \frac{4}{L^{2}}\sum_{t=1}^{\frac{L}{4}^{2}}|s_{t}|\right\rangle,
\end{equation}

where the square lattice has been divided into $\frac{L^{2}}{4}$squares
numerated by $t$ where each consists of four elementary plaquettes. Here
$s_{t}$ is the sum of the phase difference around a four-plaquette
$s_{t}\equiv(1/\pi)\sum_{\langle ij\rangle\in t}\phi_{ij}$ which means that
$|s_{t}|$ can be $0,1$ or $2.$ Thus the defect density can be described in the following way: Think of the elementary plaquettes as being either black ($s=1$) or white ($s=-1$). There are always equally many black and white squares. The defect density measures the average difference in the number of white and black squares contained in a four-plaquette. Obviously the checkerboard groundstate corresponds to a zero defect density $n_{k}=0$. However, for a finite temperature the checkerboard groundstate may contain a kink. This situation is illustrated in Fig.\ 4: start
from a checkerboard pattern. The thick dotted line is a boundary between the
two possible checkerboard patterns. The 90 degree turn of this line is
associated with a four-plaquette with $s_{t}=1$ which is denoted as thick solid 
line surrounding four plaquettes in Fig.\ 4. Thus a kink corresponds to a defect with $|s|=1$ according to our definition. The defect density defined here can be regarded as a generalization of the kink concept, since it does not rest on the possibility of uniquely identifying domain boundaries. Thus the defect density remains a well defined concept even when the checkerboard symmetry is completely restored and $m=0$. The groundstate shown in Fig.\ 2b is an example of a situation when $m=0$, because switching between + and - in Fig.\ 2b does not involve passing any energy barrier. Thus the defect density remains finite as $T$ is lowered towards zero for any $p> p_c$. Consequently, the groundstate in Fig.\ 2b corresponds to a finite defect density $n_{k}>0$. It is also obvious that the defect density is monotonously increasing with $T$.

\begin{figure}[ptb]
\begin{center}
\includegraphics[width=0.6\columnwidth]{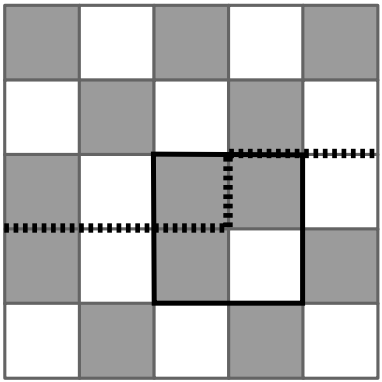}
\caption{Two checkerboard states with the boundary between them (denoted as
thick dotted line). A kink exists where the boundary makes a
90-degree turn, and the kink density $n_k$ is measured by Eq. (5).
For the four plaquettes surrounded by thick full line, $|s_t| = 1$ whereas
all other four plaquettes have even number of vortices and thus $|s_t| = 0$.}
\end{center}
\end{figure}

A phase transition associated with this order parameter is signaled by either a discontinuous or a non-analytical behavior of $n_{k}$ as a 
function of $T$ and $p$. The defect density makes a ddiscontinuousjump from zero to a finite value at $p_{c}$ in the limit of small temperatures and these two values are separated by an infinite energy barrier: the point $(p,T)=(p_{c},0)$ is the starting point of a phase transition line (see Fig.3). On this phase transition line the order parameter $n_{k}$ can only take on two values. These two values are equally probable but are separated by an infinite free energy barrier. Thus the order parameter $n_{k}$ on this phase transition line possesses a $Z_{2}$ symmetry which is broken.
 
One should note that in case of $n_{k}$ the infinite free energy barrier between two different but equally probable values of $n_{k}$ only resides on well defined lines in the $(p,T)$-plane, whereas the infinite barrier for the chirality transition resides on an area of the $(p,T)$-plane (see Fig.3). Thus the phase transition associated with the defect density $n_{k}$ is more akin to a liquid-gas transition in the pressure temperature plane: the order parameter is the density difference on the two sides of the transition line and the infinite free energy barrier only exists precisely on the transition line.

The $\mathcal{U}(1)$-symmetry is in 2D at most ``quasi''
broken because of the Mermin-Wagner Theorem\cite{mermin66}. As a consequence the corresponding phase transitions cannot be described by a local order parameter. Instead the phase transitions can be monitored by the increase of the free energy caused by a uniform
twist $\delta$ of the spin angles across the system. Expanding the free energy
$F(\delta)$ for small values of $\delta$ to lowest orders gives%

\begin{equation}
F(\delta)=Y\frac{\delta^{2}}{2}+Y_{4}\frac{\delta^{4}}{4!}.
\end{equation}
Here, $Y$ is the helicity modulus. It is finite in the low-temperature phase
and zero in the high-temperature phase.\cite{minnhagen87} $Y_{4}$ is the fourth
order modulus and can be used to verify that the helicity modulus $Y$ makes a
discontinuous jump to zero at the transition.\cite{minnhagen03} This
discontinuous jump is a key characteristics of the
KT-transition.\cite{nelson77}$^,$\cite{minnhagen81}

\section{Phase Diagram and Phase Transitions}

 In Ref.\ \cite{minnhagen07} the phase
transitions associated with the $\mathcal{U}(1)$-symmetry and the 
$Z_{2}$-chirality symmetry were investigated. The corresponding phase
diagram is reproduced in Fig.\ 3. This phase diagram has four sectors
corresponding to all four possible combinations of transitions for a
combined symmetry $\mathcal{U}(1)\otimes Z_{2}$: The four sectors are
characterized by the four possible combinations $(\Upsilon,m)=(0,0),(0,\neq
0),(\neq0,0),(\neq0,\neq0)$. The dashed horizontal line at $p=1$ in Fig.\ 3
corresponds to the usual FFXY model. In this case the phase $(\Upsilon
\neq0,m=0)$ is not realized \cite{minnhagen07}.

In the present paper we use all the three order parameters described in the previous section
together with Monte Carlo simulations in order to deduce the nature of the various phase boundaries.

\begin{figure}[ptb]
\begin{center}
\includegraphics[width=\columnwidth]{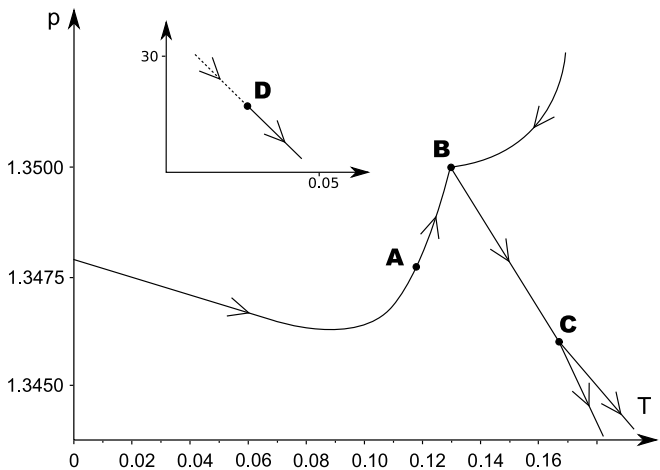}
\caption{Magnified phase diagram near $p_c \approx 1.3479$ (compare with Fig.\ 5).
There are in total four multicritical points termed A, B, C, D(see text). The critical point D shown in the inset occurs at much higher $p$ and lower $T$}
\end{center}
\end{figure}

Fig.\ 5 gives a sketch of the resulting "horizontal" phase boundary in Fig.\ 3. In this
blown up scale one finds that it has one maximum and one minimum, as well as
three multicritical points ending three distinct phase lines. The critical points
are denoted by $A,B$, and $C$. A fourth multicritical point is found along the
"vertical" phase line at much higher $p$ and lower $T$ (see Figs.3 and 5).
Let us first consider the phase boundary from $T=0$ to the critical point $A$. 
Across this first section of the phase boundary the phase transition associated with the defect density $n_{k}$ is first order.
Fig.\ 6a illustrates the discontinuous
change in the defect density $n_{k}$. The
defect-density histogram along this phase line has two distinct values of equal
probability which remain distinct in the large $L$-limit. An example is given
in Fig.\ 6b: For a given temperature $T$, the lower value corresponds to the
low-$p$ phase and the higher to the high-$p$ phase. As pointed out above, this is aanalogousto the
density for a liquid-gas transition. Note that for $T=0.1$ the $p$-value for
the first order line is lower than $p_{c}(0)$. However, as $T$ is further increased, 
the $p$-value for the first order line increases. Finally, at a
critical temperature $T_{cA}$ the density difference vanishes with increasing 
system size. This is the signature of the critical point $A$ which is hence the critical point ending the first order transition line for the defect-density. Thus the critical point $A$ is analogous to the critical point ending the first order line for a gas-liquid transition. 
Fig.\ 6c shows the defect-density histogram close to the
critical point: at the critical point the free energy barrier between the two
phases is $L$-independent. To good approximation this means that the ratio
between the maximum and minimum in the kink-density histogram should be
size-independent, whereas it increases (decreases) for lower (higher)
temperatures \cite{binder92}. This condition is fulfilled to good approximation
for the $T$-value in Fig.\ 6c. At the critical point the defect-density difference
$\Delta n_{k}$ (the difference between the two maxima in Fig.\ 6c) should vanish
with size as $\Delta n_{k}\sim L^{-\beta/\nu}$ \cite{binder92}. The $\Delta
n_{k}$ size scaling is shown in Fig.\ 6d and is consistent with an exponent
$\beta/\nu=0.25$. One can express this exponent in terms of the central charge
$c$ as $\beta/\nu=c/4$ \cite{Francesco}$^,$\cite{central}. The central charge $c$ is coupled to the symmetry of the order parameter. The defect density, the staggered magnetization and the magnetization for the 2D Ising model can all aacquireprecisely two distinct values with equal probability separated by an infinite energy barrier. The broken symmetry reflected by these order parameters does hence have a $Z_{2}$-character   and the phase transitions are Ising like. The central charge is $c=1/2$ for 2D Ising like transitions. If the order parameter on the other hand is a 2D vector then the symmetry is $U(1)$ (which means that the order parameter with equal probability have the same magnitude and any direction, but that all these possibilities are separated by an infinite energy barrier) then the central charge is $c=1$. Provided that our three order parameters covers all possibilities, then a phase transition can \emph{a priori} be any combination of single and joint transitions involving these order parameters and is hence contained within the designation $\mathcal{U}(1)\otimes Z_{2}\otimes Z_{2}$. These implies that the central charge can have the four
values $c=1/2$, $1$, $3/2$, and $2$. Here a $Z_{2}$-transition corresponds to
$c=1/2$, an individual KT-transition or a combined $Z_{2}\otimes Z_{2}$
transition corresponds to $c=1$, the two possible combined $\mathcal{U}%
(1)\otimes Z_{2}$-transitions correspond to $c=3/2$, and a combined
$\mathcal{U}(1)\otimes Z_{2}\otimes Z_{2}$ corresponds to $c=2$. These
possibilities are tested in Fig.\ 6d and singles out $c=1$ or equivalently
$\beta/\nu=1/4$. This means that of the four possible values only $c=1$ is
consistent with the data. As will be explained below, the helicity modulus
remains non-zero in this part of the phase diagram (compare Fig.3) and consequently this
suggests that the critical point $A$ reflects a combined $Z_{2}\otimes Z_{2}$
defect-density and chirality transition. The defect-density transition ends at the
critical point $A$: as $T$ is increased the free energy barrier vanishes in
the large $L$-limit. However, there is a second defect-density transition line for
higher $p$-values associated with a $\mathcal{U}(1)\otimes Z_{2}$ combined KT
and defect-density transition, as illustrated in Fig.\ 6e and f. This transition
is first order for higher $p$ and ends at a critical point $D:$ for a $T$
higher than the critical point $D$ there is no defect-density transition, just as for
the case of the critical point $A$.

\begin{figure}[ptb]
\begin{center}
\includegraphics[width=\columnwidth]{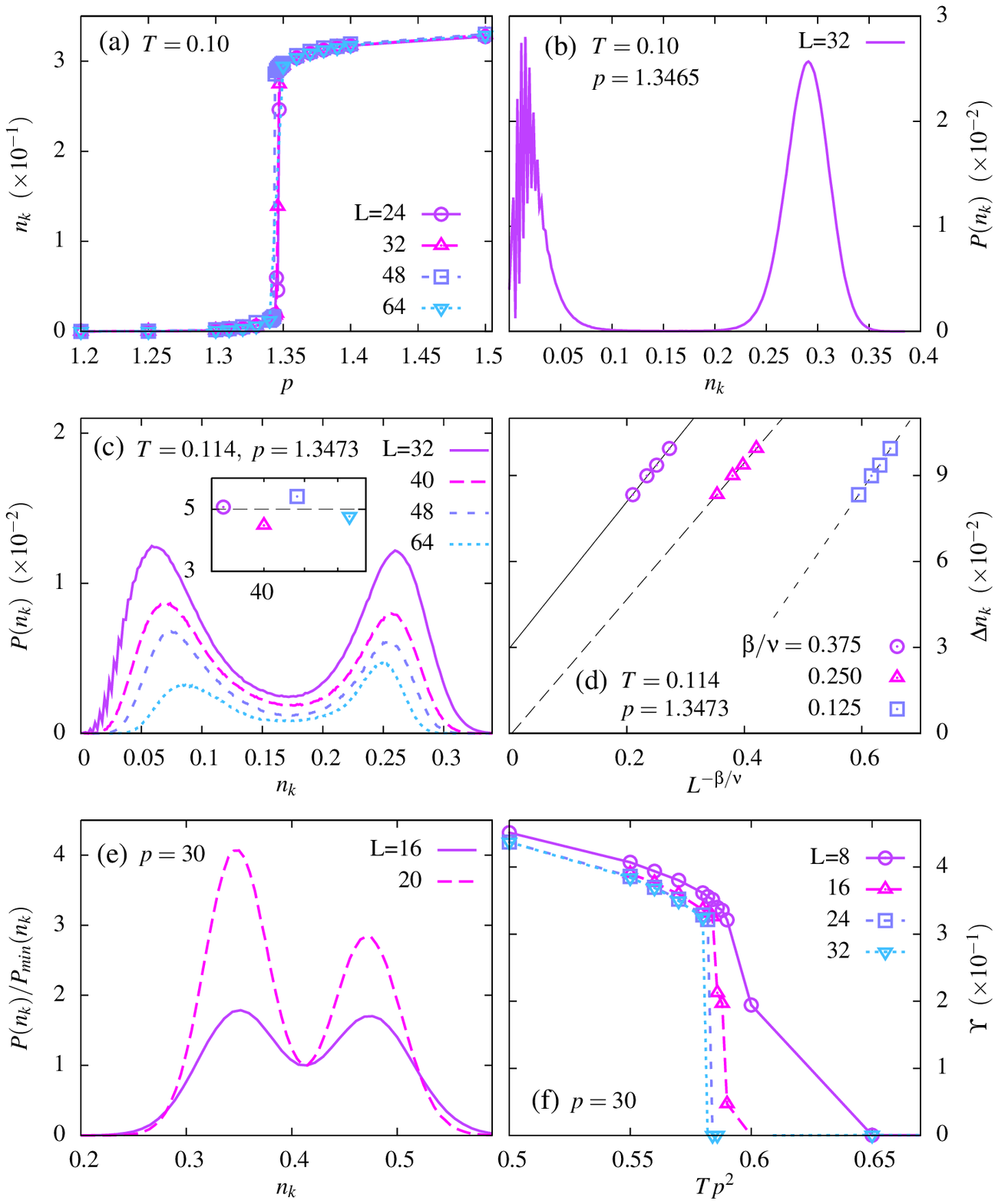}
\caption{a) First order transition of $n_k$ at $T=0.1$. b) Two valued 
probability distribution $P(n_{k})$ at the abrupt change in a). c) $P(n_{k})$ 
at the critical point A. The inset shows that the ratio $P_{\max}/P_{\min}$
is to good approximation finite and independent of $L$. d) Finite size scaling 
,$\Delta n_{k}\sim L^{-\beta/\nu}$, is consistent with
$\beta/\nu=1/4$. e) $P(n_{k})$ for a large
$p$-value above the critical point D. The figure illustrates that
$P_{\max}/P_{\min}\rightarrow\infty$ with increasing system size. e) Helicity 
modulus transition at the same $T$ and $p$ as in e) indicating a joint first order transition.
}
\end{center}
\end{figure}

\begin{figure}[ptb]
\begin{center}
\includegraphics[width=\columnwidth]{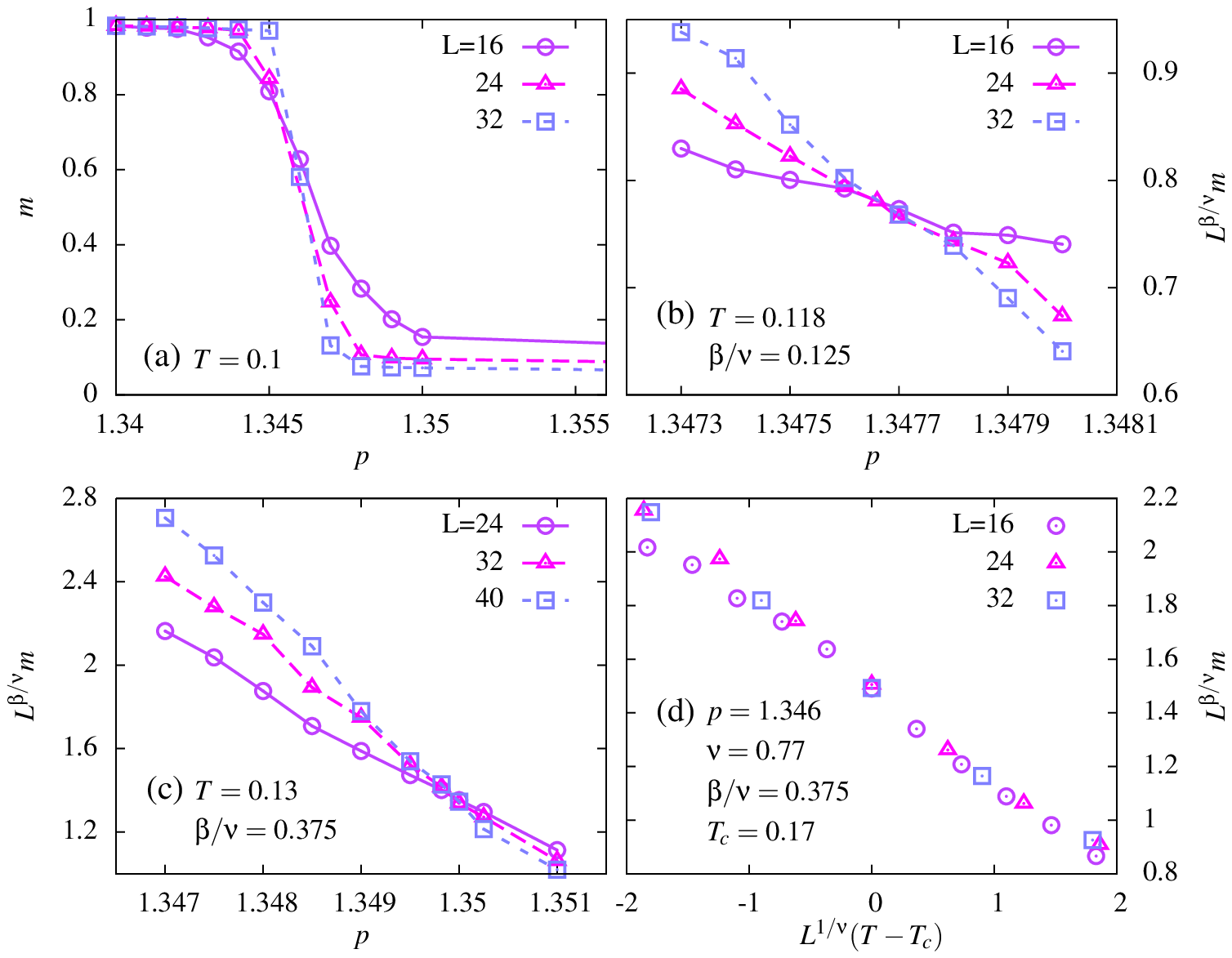}
\caption{a) The staggered magnetization $|m|$ at $T=0.1$. b)
Size scaling $|m|\sim L^{-\beta/\nu}$ for $T=0.118$ (just above point A), 
consistent with $\beta/\nu=1/8$. c) Size scaling between point B and C,
consistent with $\beta/\nu=3/8$. d) Size scaling $m=L^{-\beta/\nu}f((T-T_{cC})L^{1/\nu})$ for the
critical point C: Good scaling ccollapseobtained for $T_{cC}\approx 0.17$ with $\nu \approx 0.77$ (from Fig.\ 4(d))
and $\beta/\nu=3/8$.}
\end{center}
\end{figure}

Fig.\ 7a illustrates the chirality transition along the same phase
boundary. UUp tothe critical
point $A$ (see Fig.\ 5) the transition is first order (see 
Fig.\ 7a). The chirality transition cannot cease at the
critical point $A$ because for a fixed $T$ the free energy barrier between the
$\mathcal{O}=\pm|m|$ always vanishes for a large enough $p$. There are then two
possibilities: it can continue alone as a $Z_{2}$-transition or it can
combine with the KT-transition into a joint $\mathcal{U}(1)\otimes Z_{2}%
$-transition. To deduce which possibility is the correct one, we calculate the
size scaling of $m\sim L^{-\beta/\nu}$ and decide which of the two possible
symmetry allowed values $\beta/\nu=c/4=1/8$ or $3/8$ is consistent with the
data. Here we use standard size scaling and calculate $m(T,p)$ for a fixed $T$
for a sequence of $p$ which crosses the phase line. As seen in Fig.\ 7b, a unique
crossing point is to good approximation obtained for $\beta/\nu=1/8$. From
this we conclude that the chirality transition continues alone from the
critical point $A$ as a $Z_{2}$-transition. However as we increase the
temperature further the character of the chirality transition changes: using
the same procedure we instead find that the value $\beta/\nu=3/8$ is
consistent with the data (see Fig.\ 7c). This is consistent with
a joint $\mathcal{U}(1)\otimes Z_{2}$ KT-chirality transition. As we increase
$T$ further we come to the critical point $C$ where the KT and chirality
splits up into two separate transitions \cite{minnhagen07}. At this point it is
possible to instead calculate the size scaling for a fixed $p$. The advantage is that we can use the
standard size scaling form $m=L^{-\beta/\nu}f((T-T_{cC})L^{1/\nu}).$ This
again shows that the value $\beta/\nu=3/8$ is consistent with the data. From
this we deduce that there must exist a critical point $B$ between $A$ and $C$
where the chirality transition merges with the KT-transition.

\begin{figure}[ptb]
\begin{center}
\includegraphics[width=\columnwidth]{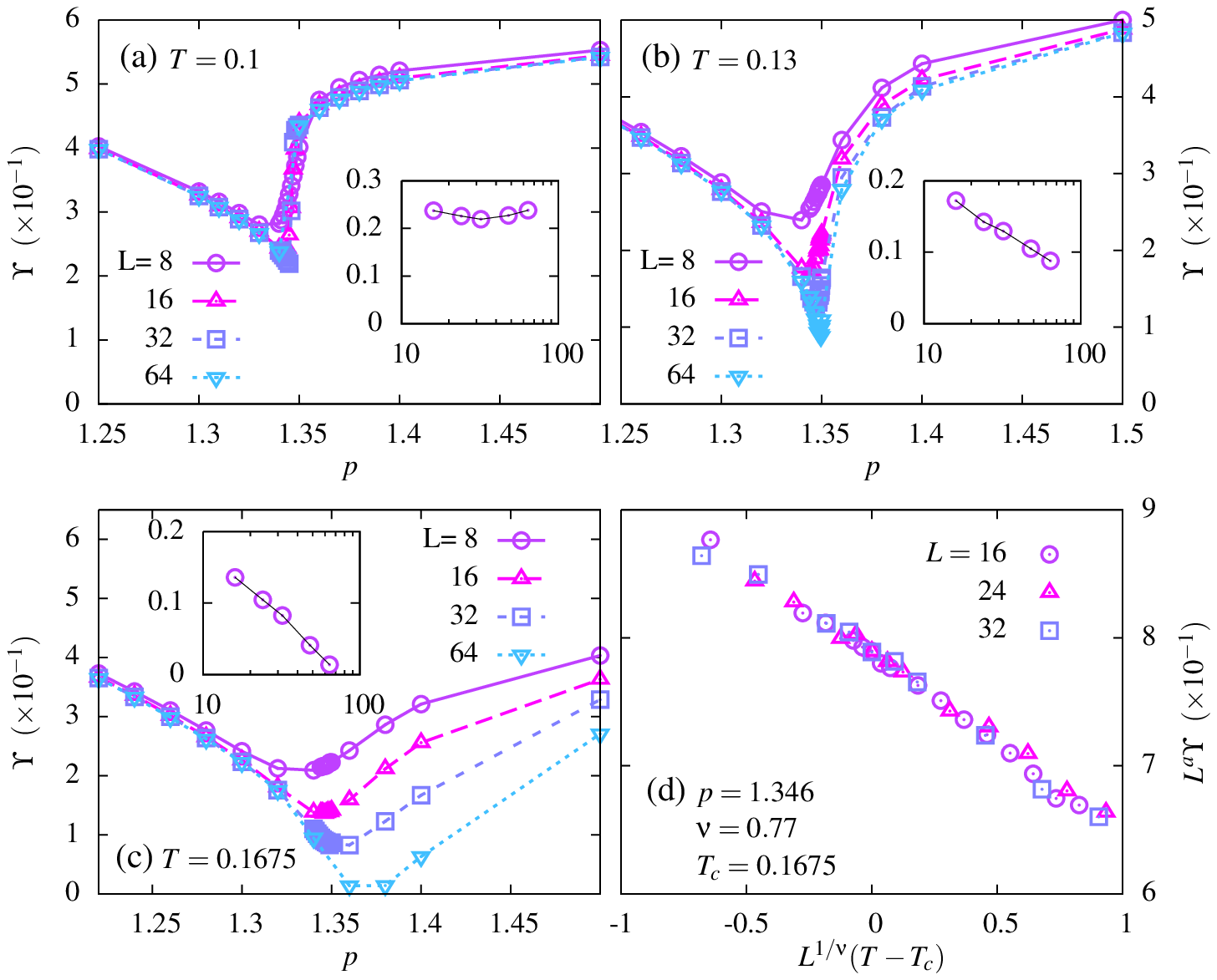}
\caption{a) The helicity modulus, $\Upsilon$, across the phase
line below point A: The inset shows that the minimum of
$\Upsilon$ remains finite with increasing size. b) $\Upsilon$ goes to
zero for large sizes between B and C. c) Same as b) at point C.
d) Size scaling relation $\Upsilon=L^{-a}g(((T-T_{cC})L^{1/\nu})$ with good ccollapse
for $\nu \approx 0.77$ ($a\approx0.63$ and
$T_{c}\approx0.1675$ are taken from Ref.\ \cite{minnhagen07}
}
\end{center}
\end{figure}

Are these deductions consistent with the $\mathcal{U}(1)$-symmetry and the
helicity modulus? We argued above that the transition from $T=0$ to the
critical point $A$ is associated with the $Z_{2}\otimes Z_{2}$ symmetry. This
presumes that the $\mathcal{U}(1)$-symmetry remains ``quasi`` broken on both sides of
the transition, or equivalently that the helicity modulus $\Upsilon$ is finite
on both sides. This is illustrated in Fig.\ 8a: the helicity modulus $\Upsilon$
has a minimum at the phase line. However, this minimum remains non-zero in the
large $L$-limit, as illustrated by the inset in Fig.\ 8a. Thus $\Upsilon$ makes
at most a finite jump at the transition and the $\mathcal{U}(1)$-symmetry remains ''quasi'' broken. Next we argued that
between the critical point $B$ and $C$, the transition is a combined
$\mathcal{U}(1)\otimes Z_{2}$-KT-chirality transition. This means that the
helicity modulus must now vanish at the transition. This is illustrated in
Fig.\ 8b, which shows that the $\Upsilon$-minimum now vanishes in the large
$L$-limit (compare inset in Fig.\ 8b). Fig.\ 8c shows the same construction close
to the critical point $C$. The fact that $\Upsilon$ vanishes as a power law
can be verified for the critical point $C$ by instead varying $T$ for fixed
$p$. In these variables the critical point $C$ obeys a standard scaling
relation $\Upsilon=L^{-a}g(((T-T_{cC})L^{1/\nu})$ which confirms the power law
decay of $\Upsilon$, as opposed to the KT-universal jump signaling the
isolated $\mathcal{U}(1)$-transition for the XY-model (see
Fig.\ 8d) \cite{minnhagen07}. We also note that the obtained critical index
$\nu\approx0.77$ is consistent with the data for $m$ in
Fig.\ 7d. It is also possible to use the fourth order helicity modulus
$\Upsilon_{4}$ to determine the character of the $\mathcal{U}(1)$%
-transition \cite{minnhagen03}. In Ref.\ \cite{minnhagen07} it was found from the
$\Upsilon_{4}$-data that, in the interval $1.346\leq p\leq1.35$, the character
of the $\mathcal{U}(1)$-transition was consistent with a transition without a
discontinuous jump in the helicity modulus $\Upsilon.$ This is consistent with
a combined $\mathcal{U}(1)\otimes Z_{2}$-transition between the multicritical points B to C.

The following picture emerges: A and D end two first
order phase lines. A is associated with a $Z_{2}\otimes Z_{2}$-transition with
central charge $c=1$ and D with a  $\mathcal{U}(1)\otimes Z_{2}$-transition
with $c=3/2$. B and C are both associated with  $\mathcal{U}(1)\otimes Z_{2}%
$-transitions and $c=3/2$ but are not end-points of first order lines.

\section{Standard 2D FFXY model}

The usual 2D FFXY model corresponds to the $p=1$-line in Fig.\ 3. The critical
point C for the 2D FFXY class is the closest multicritical point to the actual
phase transitions of the usual 2D FFXY model (compare Fig.\ 5). The critical
point C is characterized by the critical index $\nu\approx0.77$ and the
central charge $c=1.5$. A single $Z_{2}$ transition is characterized by
$\nu=1$ and $c=0.5$. All the earlier papers, in which it was putatively concluded that
the 2D FFXY\ model has a joint transition, the apparent value of $\nu$
was in the interval $0.77<\nu<1$ (see table 1 in Ref.\ \cite{hasenbusch05r}). 
In particular in Ref.\ \cite{granato93} the values of $\nu$ and $c$ were %
independently determined and given by $\nu=0.80(4)$ and $c=1.61(3)$. \ Thus
the apparent multicritical point for the usual FFXY model appeared to have
critical properties inconsistent with a single $Z_{2}$-transition and with
critical $\nu$-values in between a single $Z_{2}$-transition and the real
$\ \mathcal{U}(1)\otimes Z_{2}$ multicritical point C for the 2D FFXY class.
Furthermore, the closeness of the $\nu$-values and $c$-values ($\nu
\approx0.77$ and $c=1.5$ for C, respectively, $\nu=0.80(4)$ and $c=1.61(3)$
obtained for the usual FFXY model in Ref.\cite{granato93}) suggests that the
putative multicritical point found for the 2D FFXY model is an artifact of the
closeness to the real critical point C for the 2D FFXY class.

The present consensus is that the 2D FFXY\ model undergoes two separate
transition, a KT\ transition at $T_{KT}$ followed by a $Z_{2}$-transition at
$T_{Z_{2}}$ with $T_{KT}<T_{Z_{2}}$.\cite{hasenbusch05r} In particular
Korshunov in Ref.\ \cite{korshunov02} has given a general argument which 
purportedly states that $T_{KT}<T_{Z_{2}}$ should be true not only for the 2D
FFXY model, but also for the 2D FFXY class studied in the present work, provided that the interaction is
such that its groundstate is the broken symmetry checkerboard state. This is
in contradiction with the existence of the multicritical point C at $p<p_{c}$ (compare Fig.5) which does
correspond to an interaction potential with a checkerboard groundstate. We suggest that the reason for
this fallacy of the argument is connected to the closeness to the ($m$,$Y$)$=$(0,$\neq$0)-phase.

The most striking feature of the phase transition for 2D FFXY model is the
closeness between $T_{KT}$ and $T_{Z_{2}}$. The phase diagram in Fig.\ 5 gives a
scenario for which this feature becomes less surprising: The point is that the
chirality transition and the KT-transition merge and cross as a function of
$p$ for the 2D GFFXY model. It then becomes more natural that, for some values
of $p$, the transitions can be extremely close. The value $p=1$, which
corresponds to the usual FFXY model happens to be such a value.

There are many other $\mathcal{U}(1)\otimes Z_{2}$-models related to the 2D FFXY model\cite{hasenbusch05r}. Although, our results only pertain to the 2D FFXY-class defined in this paper, we note that, to our knowledge, none of the phase diagrams for related models contain a crossing of the KT and an Ising-like transition. In a vast majority, the KT-transition is always at lower temperature than the Ising-like transition or possibly merged. However, in the model in Ref.\ \cite{berge86} the situation is reversed with the Ising-like transition below or merging with the KT-transition. Also in this case a crossing is lacking. Because there is no crossing it is notoriously difficult assert whether a merging takes place or whether the two transitions are only extremely close.\cite{hasenbusch05r} For example the Ising-XY model was in Refs \cite{lee91}$^,$\cite{granato91}$^,$\cite{nightingale95} found to contain such a line of merged transitions. However, more careful MC simulations in fact suggest that the transitions are extremely close but never merge along this line.\cite{hasenbusch05r} The point to note is that for our 2D GFFXY-model the transitions \emph{cross} from which directly follows that a real merging exists in this case. We believe that this crossing is intimately related to the appearance of the additional groundstate.

\section{Final Remarks}
To sum up, we have found that the description of the phase diagram for the 2D FFXY-class of models requires at least three distinct order parameters consistent with the proposed designation $\mathcal{U}(1)\otimes
Z_{2}\otimes Z_{2}$: In addition to the usual KT $\mathcal{U}(1)$ transition and
the chirality $Z_{2}$-transition, there is also a defect-density transition with Ising like $Z_{2}$-character.
Within our simple parametrization of the interaction $U(\phi)$,
we have found that all combinations of transitions can be realized except two:
the single $Z_{2}$-defect transition and the fully combined $\mathcal{U}(1)\otimes Z_{2}\otimes Z_{2}$-transition.
All the others are realized, i.e.
the single $Z_{2}$-chirality transition, the single $\mathcal{U}(1)$-KT
transition, the combined $Z_{2}$-defect and $Z_{2}$-chirality transition, the
combined $Z_{2}$-chirality and $\mathcal{U}(1)$-KT, and the combined $Z_{2}$-defect
and the $\mathcal{U}(1)$-KT transition. Since the GFFXY-model is a
subclass of the 2D FFXY-class this means that at least five of the symmetry
allowed transitions can be realized. What about the remaining two? Here we
speculate that a single $Z_{2}$ -density transition will hardly be realized
because it couples too strongly to the other transitions. However, one might
imagine that there exists a potential $U(\phi)$ for which the two nearby
critical points $A$ and $B$ are merged. This critical point would then
correspond to a merged $\mathcal{U}(1)\otimes Z_{2}\otimes Z_{2}$-transition
with central charge $c=2$.

We also note that Cristofano et al in
Ref.\ \cite{cristofano06}, argued from general symmetry considerations that the
full symmetry of the FFXY-model allows for $\mathcal{U}(1)\otimes Z_{2}\otimes
Z_{2}$. The present results for the phase diagram of the 2D GFFXY model supports this designation. 

\section*{Acknowledgments}
P.M. and S.B. acknowledge support from the Swedish Research Council grant 621-2002-4135.
BJ.K. acknowledges the support by the KRF with grant no. KRF-2005-005-J11903.

\end{document}